\documentclass[epj]{svjour}
\usepackage{graphics}
\begin{document}
\title{Short-time dynamics of a packing of polyhedral grains under horizontal vibrations}
\author{Emilien Az\'ema\inst{1} \and Farhang Radjai\inst{1}  \and Robert Peyroux\inst{1} \and Vincent Richefeu\inst{1} \and
Gilles Saussine\inst{2} }
\offprints{azema@lmgc.univ-montp2.fr}         

\institute{LMGC, CNRS - Universit\'e Montpellier II, Place Eug\`ene Bataillon, 34095 Montpellier cedex 05, France 
 \and Innovation and Research Departement of SNCF, 45 rue de Londres, 75379 PARIS Cedex 08}
\date{Received: date / Revised version: date}

\abstract{
We analyze the dynamics of a 3D granular packing composed of particles of 
irregular polyhedral shape  
confined inside a rectangular box with a retaining wall subjected to horizontal harmonic forcing. 
The simulations are performed by means of the contact dynamics method for 
a broad set of loading parameters. We explore the vibrational dynamics of the packing, the evolution of solid 
fraction and the scaling of dynamics  with the loading parameters. We show that 
the motion of the retaining wall is strongly 
anharmonic as a result of jamming and grain rearrangements.   
It is found that the mean particle displacement scales 
with inverse square of frequency, the inverse of the force amplitude and the square of gravity. 
The short-time  compaction rate grows in proportion to  frequency up 
to a characteristic frequency, corresponding to collective particle rearrangements between equilibrium states, and then it declines  in inverse proportion to frequency. 
}

\maketitle

\section{Introduction}
\label{sec1}

The dynamics of dense granular materials subjected to vibrations  
involves collective phenomena resulting from 
kinematic constraints (steric exclusions, weak spatial dimensions, \ldots) 
and  energy dissipation \cite{Aranson2006}. 
Well-known examples of the vibration-induced phenomena are  compaction, 
convective flow, size segregation and standing wave patterns at the free surface\cite{Knight1993,aoki96,Clement96,liffman97,Sano2005,ribiere2005,Ciamarra2007}. 
Three different states can be distinguished depending on the intensity and 
frequency of vibrations: 
1) Gas-like or fluidized state: The rate of energy input is such that there 
are no enduring 
contacts between particles and the material behaves as a dissipative gas \cite{jaeger96b,Brennen93,Luding95}. 
2) Solid-like state: Vibrational 
energy propagates through the network of enduring  contacts between particles and the material 
undergoes slow rearrangements and progressive compaction \cite{weathers97,Josserand2000,wassgren2002,Kudrolli2004}. 
3) Liquid-like state: Both particle migration and enduring contact networks are involved in the dynamics 
and various collective effects can be observed \cite{aoki96,ben-naim96,aoki96b,liffman97,Hunt99}.            

Vibro-compaction is the main feature of the solid-like state. 
Most investigated systems are 
unconfined granular beds (with a free surface) subjected to vertical vibrations. 
The vibrations behave as a source of randomness allowing the system to explore 
metastable configurations and to reduce its potential energy. 
There is, however, another mechanism which prevails in the case of confined granular materials. 
It is well-known that under cyclic straining, a granular material accumulates 
plastic deformation and the solid fraction tends to a   maximum  value 
depending on the material. This phenomenon is sometimes called 
"granular ratcheting" due to the irreversible character of compaction under 
cyclic loading \cite{alonzo2004,Luding2004,Vandewalle2007}. 
In both confined and unconfined geometries, the 
solid fraction evolves as a logarithmic function of the number of cycles. 
In most work reported on vibrated granular media,  
 the  collective dynamics of the particles and the influence of various parameters related to the material or 
the driving system have not been investigated in all details. Moreover, in nearly all 
studies, spherical or nearly spherical particles in 3D or disks or polygons in 2D have been 
used \cite{alonzo2004,Azema2006,Vandewalle2007}.    

In this paper, we present a  numerical investigation of the dynamics 
and short-time compaction of a  
system of irregular polyhedral particles  
confined inside a rectangular box with a retaining wall subjected to  horizontal 
harmonic loading. This system is different from nearly all experimental systems investigated under 
horizontal vibrations since the packing remains confined inside a box, so that the 
gravity plays little role during the inward motion of the retaining 
wall.  This geometry is similar to that used in various industrial  
applications such as  the casting of fresh concrete where 
efficient vibro-compaction of dry and wet granular materials represents a crucial issue \cite{Swamy1976,Khayat2001}. 
The tamping operation on railway ballast  is another  interesting case  where 
vibrating tamping bars are 
used to restore the initial geometry of the track distorted as a result of 
ballast settlement\cite{SAUSSINE2004,XIMENA2001}.  
With the increase of commercial speed, a better understanding of 
the physics of compaction is important for long time stability of ballast. 

We are interested here in the evolution of the packing in the course of 
harmonic loading, the short-time compaction (during the first cycles) and 
the scaling  of the dynamics with loading parameters.  
We used discrete-element numerical simulations by means of the 
contact dynamics  method in 3D with rigid irregular polyhedral particles 
\cite{jean92,Moreau1994,Radjai1999,DUBOIS2003,Renouf2005}. 
The system is explored for a broad set of loading parameters including 
the frequency and amplitude of the harmonic driving force. A similar study 
was recently performed in 2D with polygonal particles \cite{Azema2006}. The passage 
from 2D to 3D from a purely numerical point of view involves numerical handling of 
particles of polyhedral shape and a higher numerical efficiency making 
3D simulations over many cycles and for varying parameters possible. 
In this paper, we will revisit 
the same phenomenology as in 2D for irregular polyhedral particles. This shows 
that in granular materials the dynamics is not 
sensitive to space dimensionality although the influence of particle shape 
and the details of structural organization can only be appreciated in a 3D geometry.   
    
We begin with an introduction to 3D contact dynamics method as applied to 
polyhedral particle shapes and  the numerical procedures. Then, we present in 
three sections the dynamics of the packing, the evolution of solid 
fraction and scaling with loading parameters.

\section{Numerical method}
\label{sec2}

The simulations were carried out by means of the contact dynamics (CD) method with 
irregular polyhedral particles. In this section we present the properties of this numerical method and compare it
to a more classical numerical approach molecular dynamics (MD). Then we presents the principle of the contact detection and numerical parameters which have been used for this study.

\subsection{Contact dynamics}

The CD method is based on implicit time integration and nonsmooth formulation of
mutual exclusion and dry friction between particles in case of contact \cite{jean92,Moreau1994,Radjai1999,DUBOIS2003}.
The equations of motion for each particle are formulated as differential inclusions in which
velocity jumps replace accelerations \cite{Moreau1994}. 
The unilateral contact interactions and Coulomb friction law are represented as set-valued
force laws according to convex analysis. 
The implementation of the time-stepping scheme requires the 
geometrical description of each potential 
contact in terms of contact position and its unit normal vector.

At a given step of evolution, all kinematic constraints implied by enduring  contacts and the possible rolling of some particles over others are {\it   simultaneously} taken into account, together with  the equations of dynamics, in order to determine all velocities and contact forces in the system. 
This problem is solved by an iterative process (non-linear Gauss-Seidel method) which consists
of solving a single contact problem, with other contact forces being treated as known, and iteratively 
updating the forces until a given convergence criterion is fulfilled. 
The method is thus able to deal properly with the {\it nonlocal} character of the momemtum transfers --- resulting from the perfect rigidity of particles in contact. The CD method makes no difference between smooth evolution of a system of rigid particles during one time step and nonsmooth evolutions in time due to collisions or dry friction effects.

The  MD-like methods are based on regularization schemes where impenetrability is approximated by a steep repulsive potential and Coulomb's law by a viscous- or elastic-regularized friction law, to which smooth computation methods can be applied. In this case, the choice of viscous or elastic parameters  has to be specified depending 
on the particle shapes. This regularization implies small time steps in order to ensure numerical stability  
whereas the implicit time integration method, inherent in the CD method, is unconditionally stable. 
The uniqueness is not guaranteed by CD approach for perfectly rigid particles in absolute terms. However, by initializing each step of calculation with the forces calculated  in the preceding step, the set of admissible solutions shrinks to fluctuations which are basically below the numerical resolution. In MD-based simulations, this ``force history" is  encoded  by construction in the particle positions.

For our simulations, we used the LMGC90 which is a multipurpose software developed in Montpellier, capable of modeling a collection of deformable or undeformable particles of various shapes (spherical, polyhedral, or polygonal) by different algorithms\cite{DUBOIS2003}.

\subsection{Simulation of polyhedral particles}

The treatment of a contact interaction between two particles requires 
the identification of the contact zone and a plane (in 3D), the so-called ``common plane". 
Obviously, for rigid particles it is possible to define this contact zone by a finite set of points.
Before applying the contact detection algorithm between a pair of particles of  
irregular shapes (see Fig. \ref{fig2_grain}), a ``bounding box" method is used to compute a list 
of particle pairs potentially in contact.
Then, for each pair, the first step is to determine if an overlap exists through a 3D 
extension of the ``shadow overlap method" \cite{SAUSSINE2004,DUBOIS2003}.
 Several algorithms exist for overlap determination between convex polyhedra   \cite{nezami2004,DUBOIS2003,Saussine2006}. In the case of an overlap, the contact plane is determined by 
computing the intersection between the two particles.

\begin{figure}
\centering
\resizebox{0.2\textwidth}{!}{\includegraphics{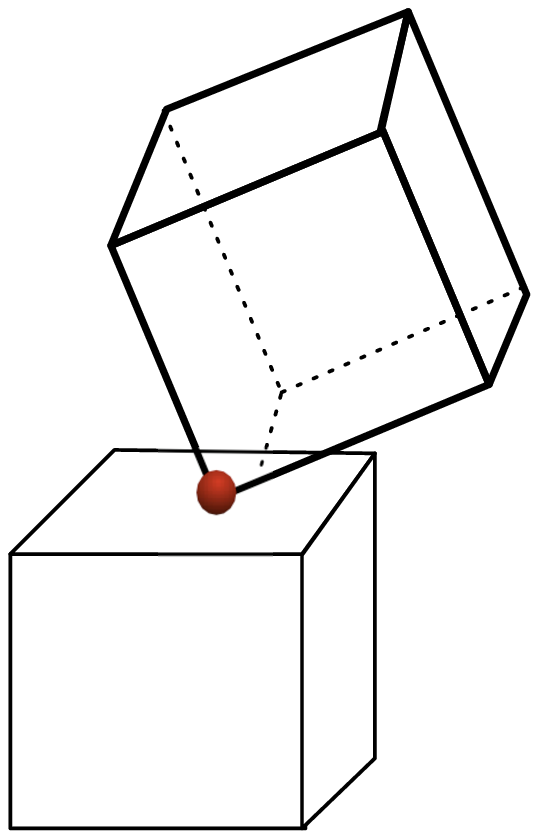}}
\resizebox{0.15\textwidth}{!}{\includegraphics{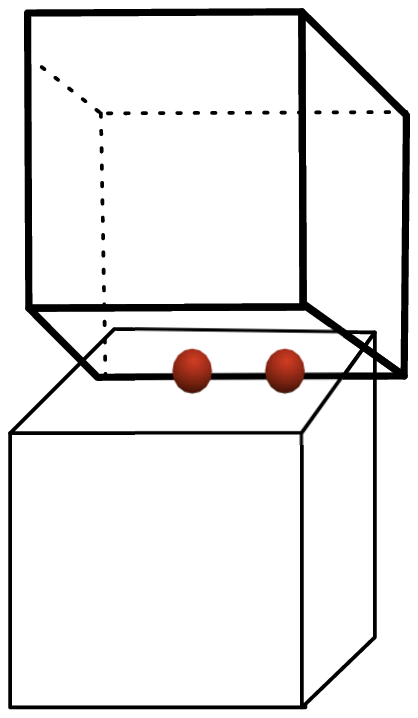}}
\resizebox{0.2\textwidth}{!}{\includegraphics{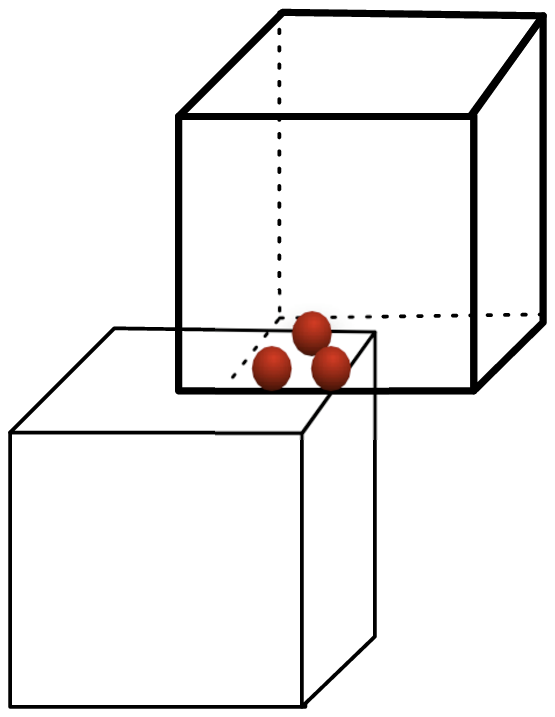}}
\caption{Different types of contacts between two polyhedra.}
\label{fig1_grain}    
\end{figure}

The contacts between polyhedral particles belong to different categories, namely face-face, edge-face, vertex-face, edge-edge, vertex-vertex, vertex-edge; see Fig. \ref{fig1_grain}. The vertex-vertex and vertex-edge 
contacts are rare, i.e. in practice of ``zero measure''. For each case we determine one, two or three contact points which provide a good description of the contact zone.
In this paper, the vertex-edge and edge-edge contacts are referred to  as ``simple" contacts. 
On the other hand, the edge-face and face-face contacts are treated as 
``double" and ``triple" contacts, respectively, as their representation 
involves 2 and 3 distinct points on the common plane. 
The detection procedure is fairly rapid and allows us to simulate 
large samples composed of polyhedral particles. 

\subsection{Numerical samples}
\begin{figure}
\centering
\resizebox{0.5\textwidth}{!}{\includegraphics{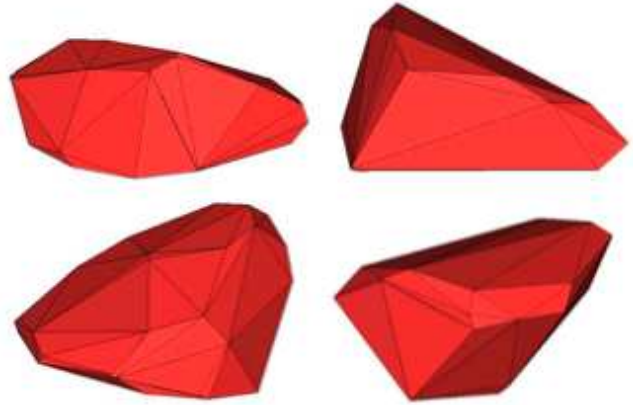}}
\caption{Examples of polyhedral shapes used in the simulations.}
\label{fig2_grain}    
\end{figure}

\begin{figure}
\centering
\resizebox{0.4\textwidth}{!}{\includegraphics{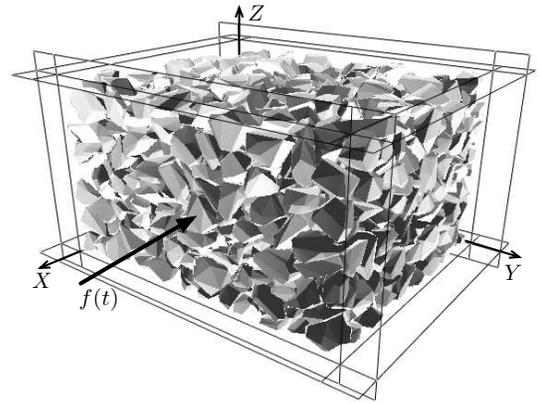}}
\caption{A snapshot of the packing inside a box with a free wall over which 
the driving force $f(t)$ is applied along the $x$ direction. }
\label{fig1}    
\end{figure}

Our numerical samples  are composed of rigid polyhedral particles with shapes 
and sizes that represent those of ballast grains (Fig. \ref{fig2_grain}). 
Each particle has at most 
70 faces and 37 vertices and at least 12 faces and 8 vertices. A sample contains nearly 
1200 particles. 
The particle size is characterized as the largest distance between the 
barycenter and the vertices of the particle, to which we will refer as ``diameter" below. 
We used  the following size distribution:  
50\% of diameter $d_{min}=2.5$ cm, 34\% of diameter $3.75$ cm, 
16\% of diameter $d_{max}=5$ cm. 
The particles are initially placed in a  
rectangular box and compressed by downward motion of the upper wall 
 at zero gravity; see Fig. \ref{fig1}. Then, the gravity is set to $g$ and the 
upper wall is raised 1 cm and fixed. The coefficient of friction between the particles  
and with the horizontal walls 
was fixed to 0.4, but it was 0 at the vertical walls. 
The coefficient of restitution between particles was fixed to zero because 
of the high solid fraction of the samples. 
One of the walls   is allowed to move 
horizontally (x direction in Fig. \ref{fig1}) and subjected to a harmonic driving force. 
All other walls are immobile.  
For all simulations the time step was $2.10^{-4}s$.

\section{Vibrational dynamics}
\label{sec3}

The free wall is subjected to a harmonic force as a function of time, 
\begin{equation}
f(t)=\frac{(f_{max}+f_{min})}{2}-\frac{(f_{max}-f_{min})}{2}\sin \omega t, 
\end{equation}
where $f_{max}$ and $f_{min}$ are the largest and lowest compressive 
(positive) forces acting on the wall. The first term represents the mean confining 
force modulated by the second term.   
At $t=0$, the external force 
$f=(f_{max}+f_{min})/2$ causes the inward motion (contraction) of the free retaining wall. 
Jamming occurs when the gap left between the upper wall 
and the free surface of the packing is filled.  
If $f_{min}$ is above the (gravitational) force 
exerted by the particles on the free  wall, $f$ will be large enough 
to prevent the wall from backward motion (extension) during the whole 
cycle.  Then, the granular material is in ``passive state"  and the 
major principal stress direction is horizontal \cite{nedderman92}. 
On the other hand, if $f_{max}$ is below the force exerted by the particles, 
$f$ will never be large enough to prevent the extension of the packing. 
This corresponds to  the ``active state" where the major 
principal stress direction remains vertical. 

In all other cases, both contraction and extension occur during each period, 
and the displacement $\Delta x$ of the free wall will be controlled by  $f_{min}$
 and $f_{max}$. Without loss of generality, we set $f_{min}=0$. 
This ensures the largest possible displacement of the wall in the active state. 
Four different values of $f_{max}$ were tested, 
ranging from $2.10^3$ N to $10^4$ N. 

\begin{figure}
\centering
\resizebox{0.4\textwidth}{!}{\includegraphics{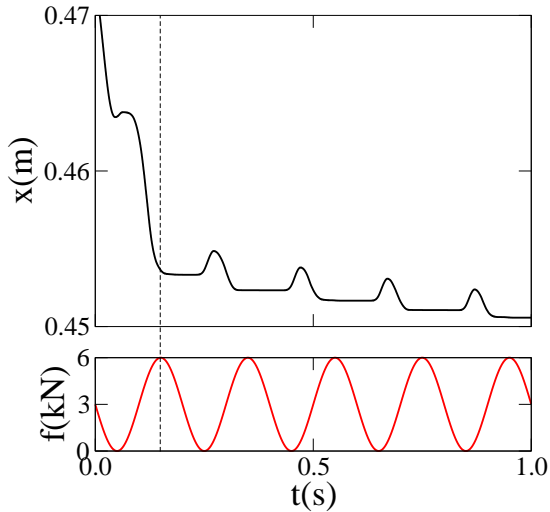}}
\caption{The evolution of the displacement $x$ of the free wall (up) in response to 
harmonic loading (down).}
\label{fig2}    
\end{figure}

We first consider the trajectory $x(t)$ of the free wall  
which reflects the dynamics of the particles in the box in 
response to harmonic forcing.
Figure \ref{fig2} shows $x(t)$  
for frequency $\nu=5$ Hz over a time interval $\Delta t = 1$ s. 
We can observe a fast initial contraction ($t<0.1$ s) followed by slow 
contraction over four periods. 
The initial contraction is a consequence of the gap left 
between the free surface of the packing and the upper wall. 
The subsequent periodic motion of the wall takes place around this 
jammed state and will be studied below.

\begin{figure}
\centering
\resizebox{0.4\textwidth}{!}{\includegraphics{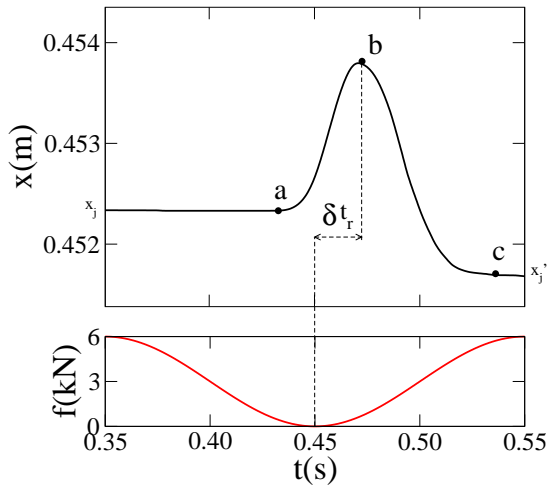}}
\caption{A zoom on the displacement of the free wall as a function of time for a single period (up) 
in response to harmonic force (down).}
\label{fig3}    
\end{figure}

\begin{figure}
\centering
\resizebox{0.4\textwidth}{!}{\includegraphics{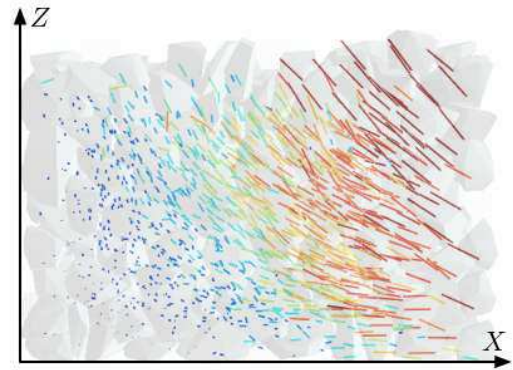}}
\caption{Instantaneous particle velocity field  in the passive state, i.e. during 
inward motion of the free wall.}
\label{fig6}    
\end{figure}

A zoom on a single period is shown in Fig. \ref{fig3}.
It begins at the jamming position $x=x_j$ corresponding to 
the jamming position reached at the end of the preceding period. 
The motion of the wall begins (point a in Fig. \ref{fig3}) only when the 
applied force $f$ declines near to its minimum $f_{min}=0$. 
The maximum displacement $\Delta x_{max}$ occurs at a later time $\delta t_r$ (point b). 
From a to b, the  force exerted by the packing on the free wall is above 
the applied force, so that the wall moves backward (extension). 
In this phase, the packing is in an active state. 
The inverse situation prevails from b to c where the particles are pushed towards the box (contraction). 
Then, the packing is in a passive state. 
The new jamming position $x'_j$ is below the jamming position $x_j$ 
reached at the end of the preceding period. 
The difference $x_j - x'_j$ represents the net compaction of the packing 
over one period. 
The particle velocity field is not a simple oscillation around an average position. 
The particles undergo a clockwise convective motion in the cell as shown in Fig. \ref{fig6}.  

\begin{figure}
\centering
\resizebox{0.4\textwidth}{!}{\includegraphics{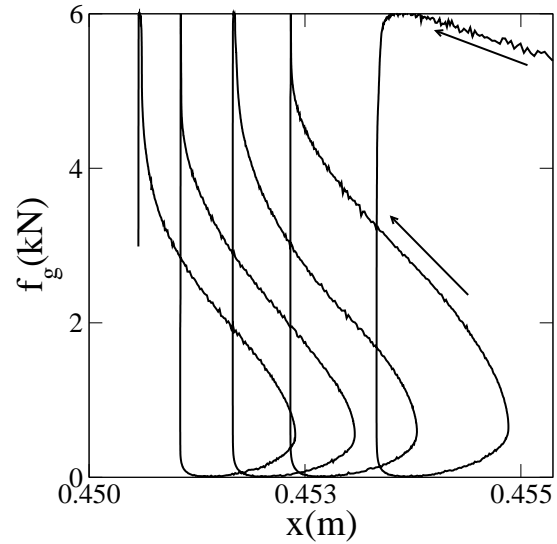}}
\caption{Force $f_g$ exerted by the particles on the free wall 
as a function of displacement $x$.}
\label{fig5}    
\end{figure}

\begin{figure}
\centering
\resizebox{0.4\textwidth}{!}{\includegraphics{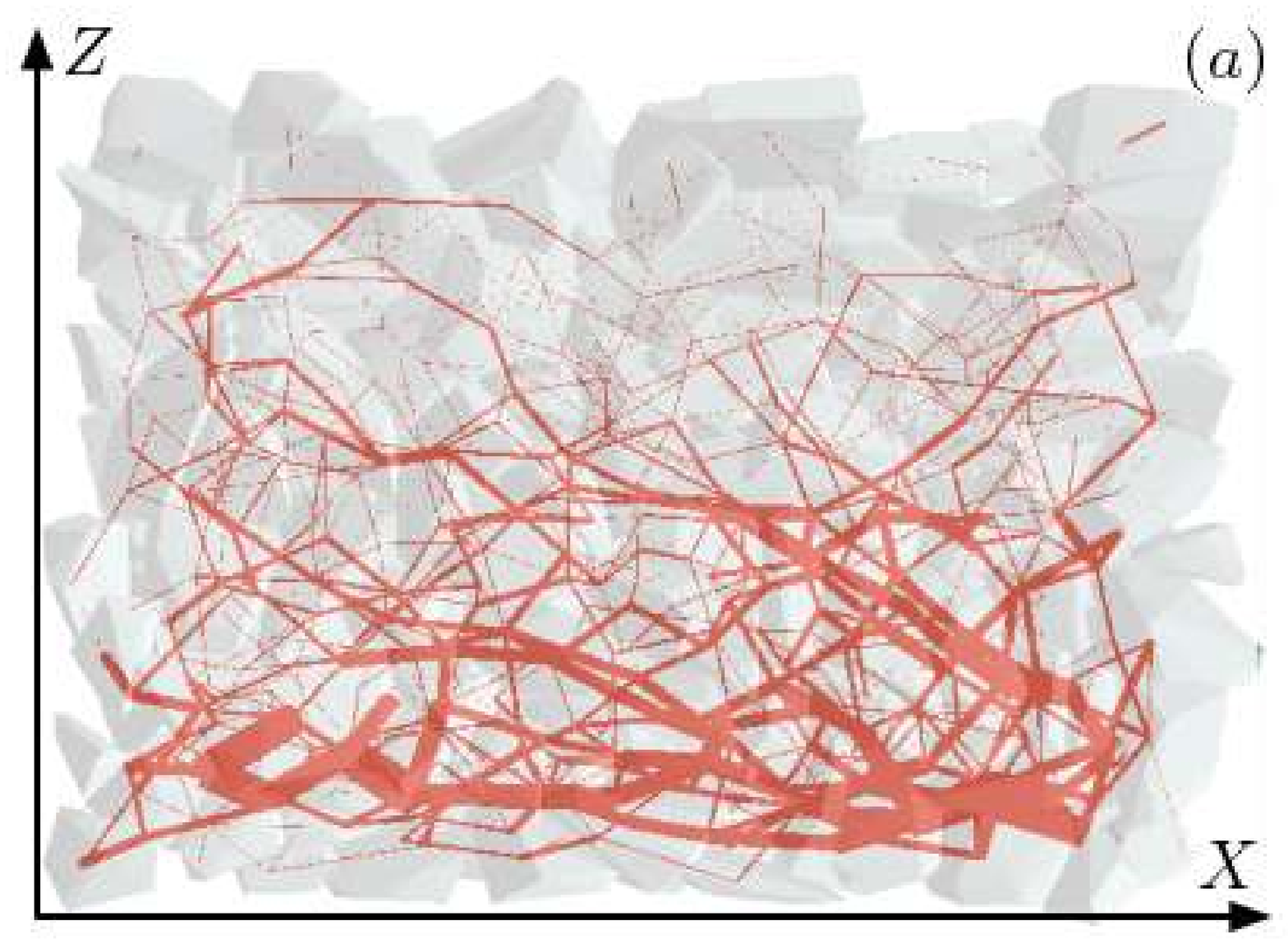}}
\resizebox{0.4\textwidth}{!}{\includegraphics{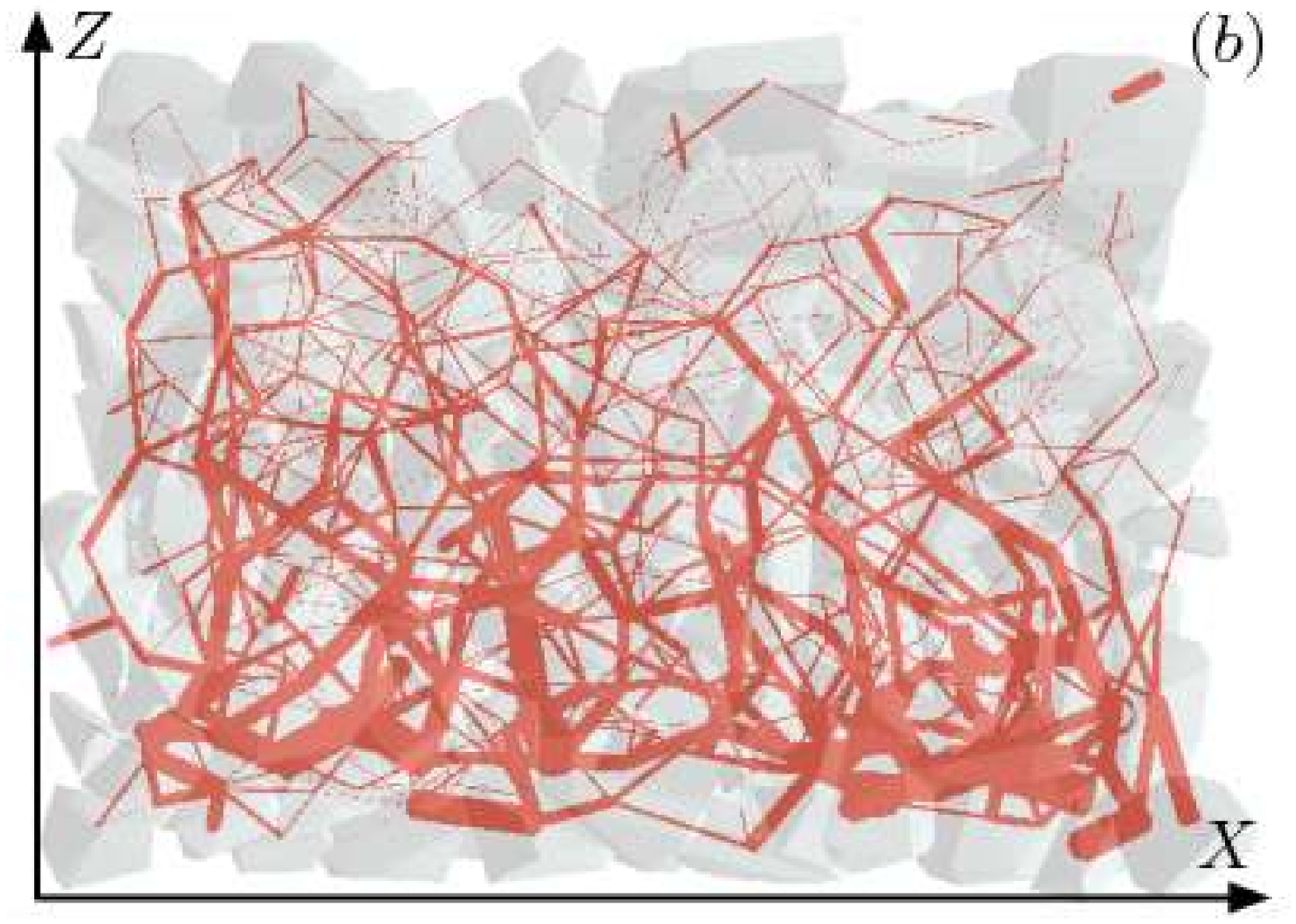}}
\caption{Normal forces in the passive (a) and active (b) states in a section of the 
packing parallel to the xz plane. 
The segments connect particle centers with a  thickness proportional to the normal force.}
\label{fig7}    
\end{figure}

Figure \ref{fig5} shows the horizontal force $f_g$ exerted by the packing on 
the wall as a function of $x$ over four periods. 
In the active phase, $f_g$ grows slightly with $x$.  
In the passive phase, it grows faster and almost linearly as $x$ decreases. 
The vertical line corresponds to the jammed state 
where $f_g$ decreases with $f$ at $x=x_j$. 
We also observe   two transients :
1) unjamming and the onset of the active state, 
2) jamming from the passive state. 
Inside the packing, the contact forces evolve between a fully jammed state, 
where horizontal force chains dominate (Fig. \ref{fig7}(a)), and the active state, 
where vertical gravity-induced chains can be observed (Fig.  \ref{fig7}(b)).

\section{Granular ratcheting}
\label{sec5}

\begin{figure}
\centering
\resizebox{0.4\textwidth}{!}{\includegraphics{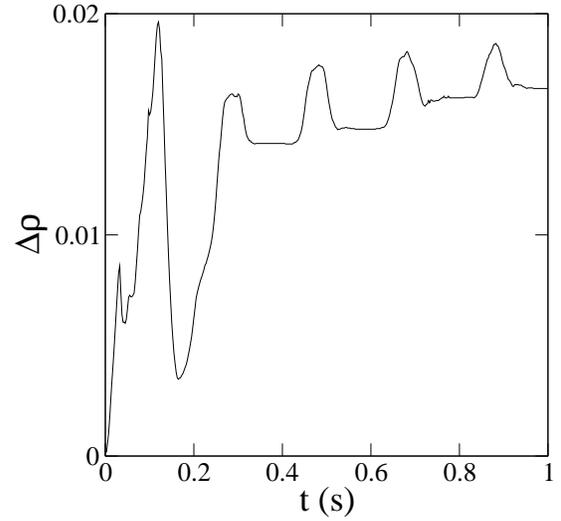}}
\caption{Evolution of the solid fraction $\Delta\rho$ from the initial state 
as a function of time over several periods.}
\label{fig11}    
\end{figure}

In our system, the solid fraction $\rho$ of the packing increases due to horizontal vibrations. 
This accumulation of plastic strain under oscillatory loading is sometimes 
called ``granular ratcheting" \cite{alonzo2004,Luding2004}. 
To  evaluate $\rho$, we consider a control volume  inside the box. 
The initial value of the solid fraction is $0.50$.   
Figure \ref{fig11} shows the evolution of the variation $\Delta\rho$ of solid 
fraction for several periods. An initial compaction of $2\% $ is followed 
by oscillations  with a small increase of $\Delta\rho$ in each period. 
The initial compaction should be attributed to the initial state of 
the packing is not yet fully confined. 
We use $\rho_0 = 0.51$, reached after a time laps 
of $0.2$ s, as the reference 
value for the evolution of solid fraction. The relative compaction of the packing is given 
by $\Delta\rho / \rho_0$. The compaction rate ${\dot \eta}$ over several periods 
and for a total time interval $\Delta t$ is 
\begin{equation}
 {\dot \eta} \equiv \frac{1}{\rho_0} \frac{\Delta\rho}{\Delta t}.
 \label{eq:eta} 
\end{equation}
The compaction of the packing slows down logarithmically at  long times \cite{ben-naim96}. 
But, the short-time compaction can well be approximated by a linear function with a constant 
compaction per period  $\Delta\rho_1$. Then, we have 
\begin{equation}
 {\dot \eta}  = \frac{\Delta\rho_1}{\rho_0} \ \nu. 
 \label{eq:eta1} 
\end{equation}
For $\nu = 5$ Hz and $f_{max} = 6\;10^3$ N, we have 
${\dot \eta} \simeq 0.025$  s$^{-1}$. This rate is faster in 3D compared to 2D simulations for 
the same frequency \cite{Azema2006}.   
  
\begin{figure}
\centering
\resizebox{0.4\textwidth}{!}{\includegraphics{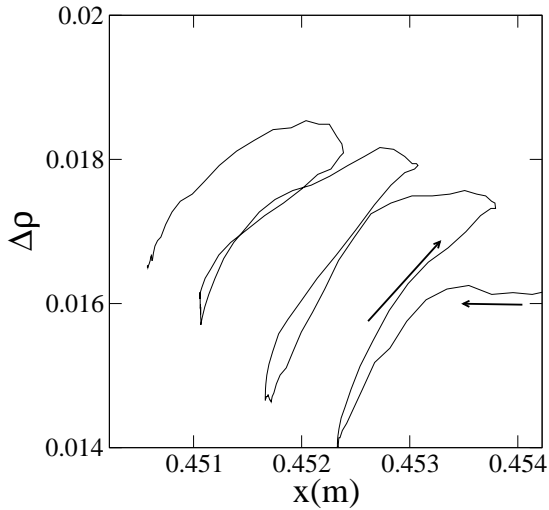}}
\caption{Variation  $\Delta\rho$ of the solid fraction from the initial state 
as a function of the displacement $x$ of the free wall.}
\label{fig12}    
\end{figure}

It is important to note that compaction occurs in the active state, i.e. during the 
extension of the packing. 
This is shown in Fig. \ref{fig12}, where the variation  $\Delta \rho$ of the 
solid fraction is plotted as a function of $x$. The solid fraction increases during extension (increasing $x$) 
and decreases during contraction (decreasing $x$). 

Granular ratcheting has been investigated by numerical simulations showing that  
the anisotropy of critical contacts, where the friction force is 
fully mobilized, plays an important role \cite{alonzo2004,Luding2004}. 
Quasi-static cyclic shearing also leads to cumulative compaction of a granular 
material at low strain amplitudes \cite{radjai01,mitchel05}. 
At large amplitudes, the compaction is followed by decompaction (dilation) 
and no net compaction can be observed over a full cycle.  
In our system,     
compaction is a consequence of unjamming and it is pursued during 
the whole active state. 
Decompaction takes place in the passive state, but it is cut short by fast jamming.  
The outcome of a full cycle is thus a net compaction of the packing. 

\section{Influence of loading parameters}
\label{sec6}

We  performed a series of simulations for frequencies $\nu$  
ranging from $1$ Hz to $60$ Hz and for a total time of 1 s.  
All simulations yield similar results both for dynamics and compaction.
Moreover, a simple dimensional analysis leads to the collapse of 
the data on a single plot. Indeed, the frequency  
sets the time scale $\tau = \nu^{-1}$. 
Force scales are set by the largest driving force $f_{max}$ in the passive state and 
the particle weights $mg$ as well as the smallest driving force $f_{min}$ in the active state. 
Hence, dimensionally, for fixed values of  $mg$, $f_{min}$ and $f_{max}$,  
all displacements are expected to scale with $\nu^{-2}$ and 
all velocities with $\nu^{-1}$. 

\begin{figure}
\centering
\resizebox{0.4\textwidth}{!}{\includegraphics{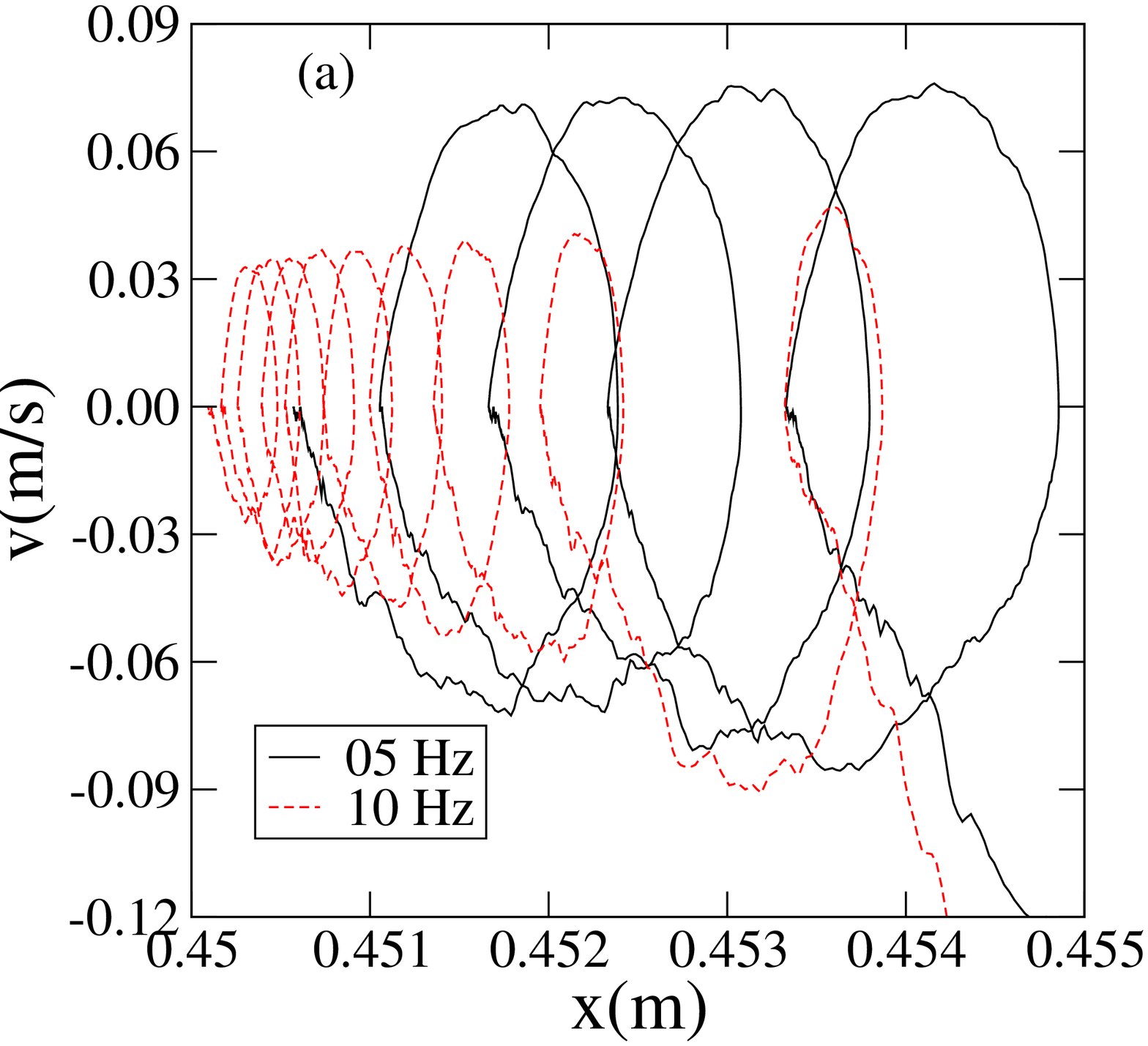}}
\resizebox{0.4\textwidth}{!}{\includegraphics{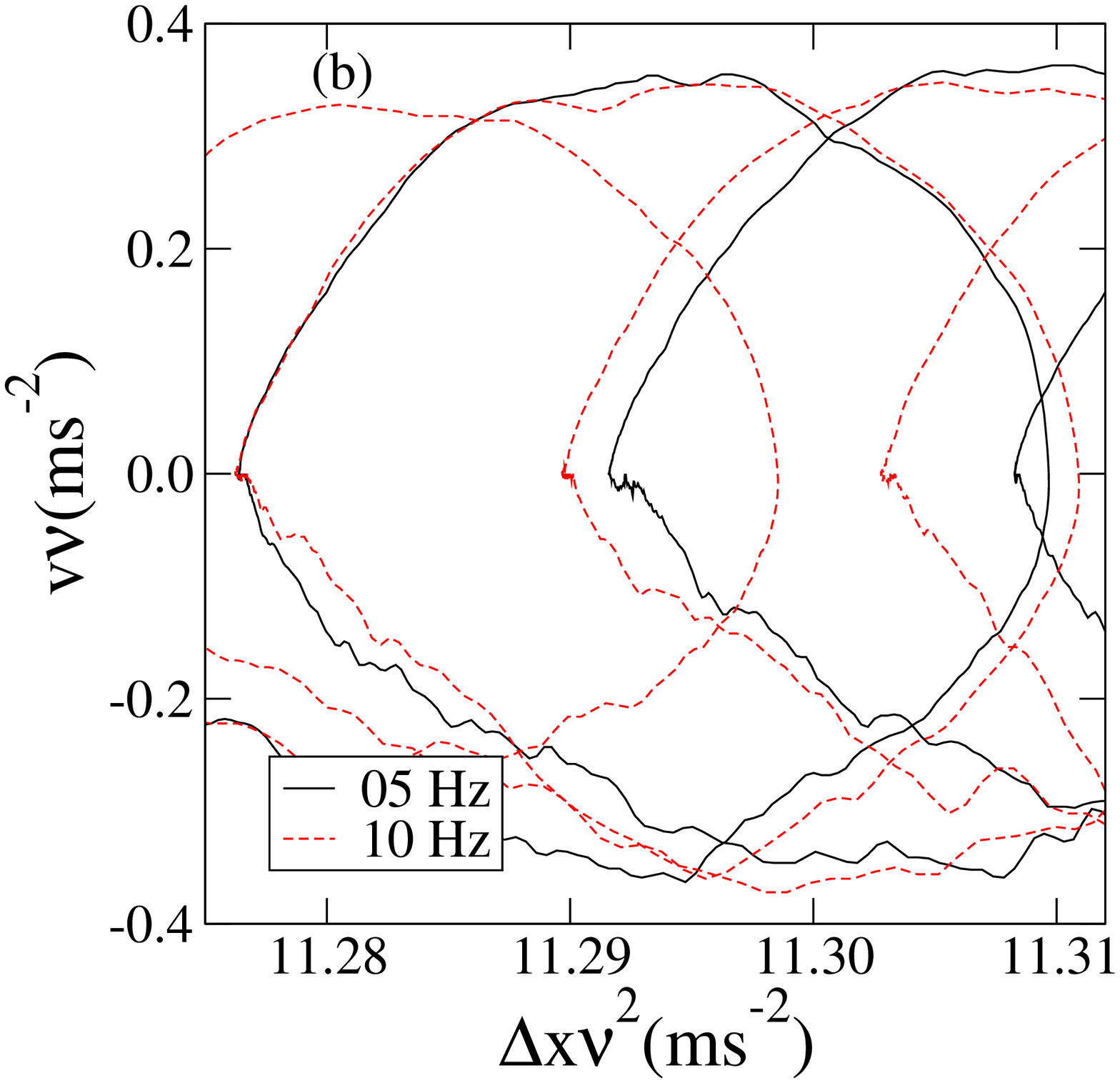}}
\caption{Phase space trajectories for two frequencies without scaling (a) 
and with scaling (b) of the displacements and velocities with respect to the frequency.}
\label{fig13}    
\end{figure}

\begin{figure}
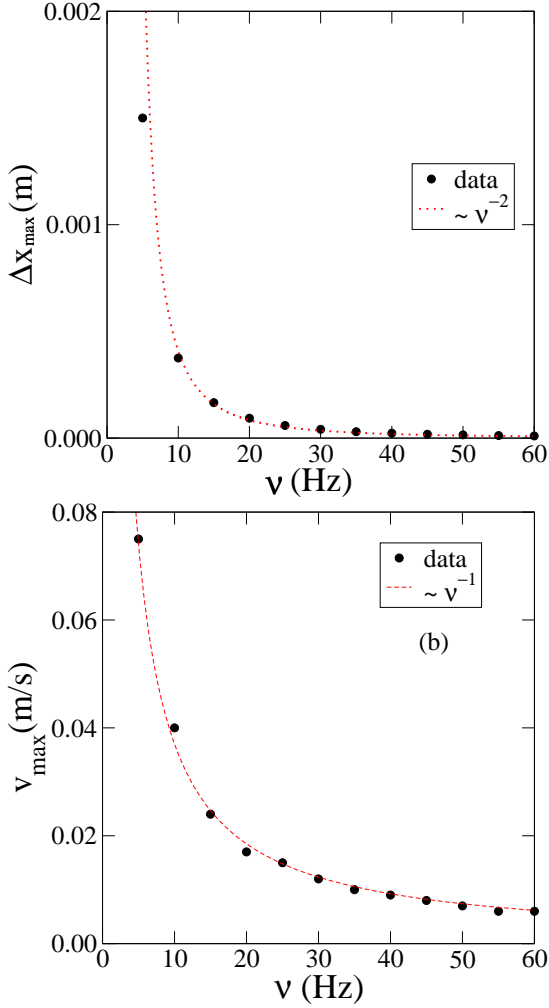

\centering
\resizebox{0.4\textwidth}{!}{\includegraphics{fig12a.eps}}
\resizebox{0.4\textwidth}{!}{\includegraphics{fig12b.eps}}
\caption{Maximum displacement ${\Delta x}_{max}$ (a) and the maximum velocity $v_{max}$ (b) 
as a function of frequency $\nu$.}
\label{fig14}    
\end{figure}

This scaling is shown in Fig. \ref{fig13} where the phase 
space trajectory is shown for $\nu = 5$ Hz and $\nu = 10$ Hz without scaling 
and  after scaling the displacements $\Delta x$ by $\nu^{-2}$ and the velocities $v$ by  $\nu^{-1}$.
We see that the data  from both simulations collapse nicely on 
the same trajectory after scaling.  Figure \ref{fig14} shows the maximum 
displacement ${\Delta x}_{max}$ in the active state and the maximum velocity $v_{max}$ in 
the passive state as a function of $\nu$. 
The corresponding fits by $\nu^{-2}$ and $\nu^{-1}$ are excellent. 
  
\begin{figure}
\centering
\resizebox{0.4\textwidth}{!}{\includegraphics{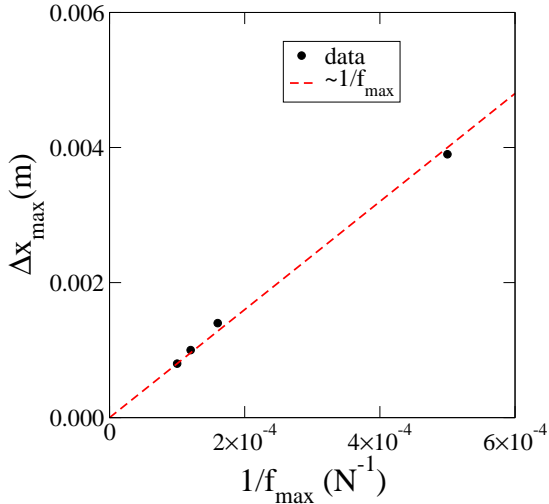}}
\caption{Scaling of the maximum displacement ${\Delta x}_{max}$ with the 
force amplitude $f_{max}$.}
\label{fig15}    
\end{figure}

The role of force parameters $mg$, $f_{min}$ and $f_{max}$ 
is less evident. Since we have $f_{min}=0$, we expect 
${\Delta x}_{max}$ to be dependent on the 
ratio $mg / f_{max}$ representing the relative importance of the gravitational  
to driving forces. Indeed, our data show that  ${\Delta x}_{max}$  
varies as $f_{max}^{-1}$; Fig. \ref{fig15}. 
On the other hand, the mass ratio $m_w/ m$, where $m_w$ and $m$ are 
the mass of the free wall and the total mass of the particles, must control the 
inertia and thus the maximum displacement of the wall. 
Our simulations with different values of $m_w$ show 
that  ${\Delta x}_{max}$  varies as 
$m/(m+m_w)$. 

\begin{figure}
\centering
\resizebox{0.4\textwidth}{!}{\includegraphics{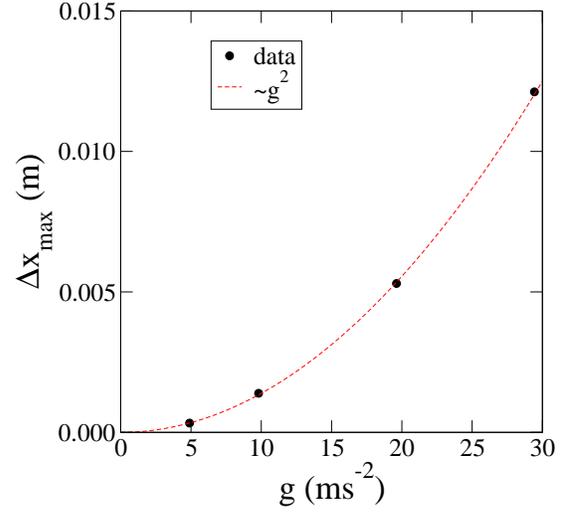}}
\caption{Scaling of the maximum displacement ${\Delta x}_{max}$ with 
gravity $g$.}
\label{fig16}    
\end{figure}

\begin{figure}
\centering
\resizebox{0.4\textwidth}{!}{\includegraphics{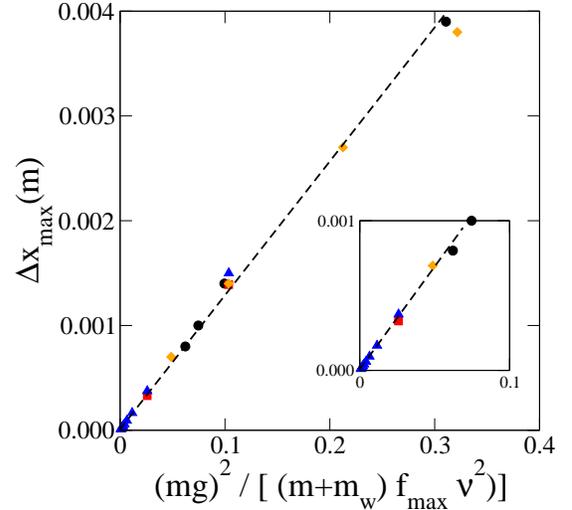}}
\caption{Scaling of the maximum displacement ${\Delta x}_{max}$ with loading parameters 
from simulations with different values of the frequency $\nu$ (triangles), the 
force amplitude $f_{max}$ (circles), the gravity $g$ (squares), and for the mass $m_w$ (diamonds) of the free wall.
The inset shows the plot near the origin. }
\label{fig17}    
\end{figure}

Hence, we propose the following expression  for the scaling of  displacements 
with loading parameters:    
\begin{equation}
 {\Delta x}_{max}  = C \left( \frac{m}{m+m_w} \right) \left( \frac{mg}{f_{max}} \right)  \left( \frac{g}{\nu^2} \right),
 \label{eq:dx}
 \end{equation}
where $C$ is a dimensionless prefactor.  
Fig. \ref{fig17} shows  
${\Delta x}_{max}$ as a function of $(mg)^2 / [(m+m_w)(f_{max} \nu^2)]$ from 
 different simulations with different values of $\nu$, $f_{max}$, $g$ and $m_w$. 
We see that the data are in excellent agreement with Eq. \ref{eq:dx}. The prefactor is  
 $C \simeq 0.01$. This scaling is the same as in 2D simulations with a material constant 
 $C \simeq 0.05$ for polygonal particles \cite{Azema2006}.  Let us also remark that 
 Eq. \ref{eq:dx} predicts that ${\Delta x}_{max}$ varies as $g^2$. 
This prediction agrees well with our simulation data 
shown in Fig. \ref{fig16} for four different values of $g$. 
 
\section{Compaction rates}
\label{sec7}

\begin{figure}
\centering
\resizebox{0.4\textwidth}{!}{\includegraphics{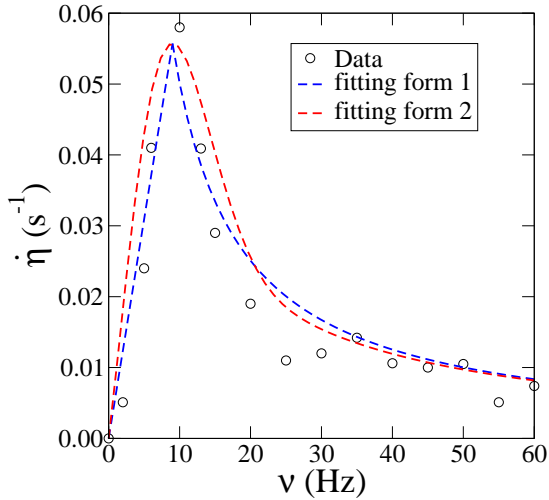}}
\caption{The compaction rate ${\dot \eta}$  as a function of  the frequency (circles) fitted by two 
different functions; see text.}
\label{fig18}    
\end{figure}

We now come back to granular ratcheting and we would like to 
evaluate short-time compaction rates as a function of frequency.  
According to Eq. \ref{eq:eta1}, the compaction rate ${\dot \eta}$ varies  
linearly with the frequency $\nu$ if the total compaction 
per period $\Delta \rho_1$ is independent of $\nu$. 
Fig. \ref{fig18} shows ${\dot \eta}$  as a function of $\nu$. 
We see that only at low frequencies, ${\dot \eta}$ increases linearly with $\nu$. 
Beyond a characteristic frequency $\nu_c$, ${\dot \eta}$ declines with $\nu$. 
The largest compaction rate ${\dot \eta}_{max}$ occurs for $\nu = \nu_c\simeq 10$ Hz. 
The corresponding time $\tau_c \equiv \nu_c^{-1}$ represents the characteristic time 
for the relaxation of the packing. 
In the active state, the packing needs a finite rearrangement time 
to achieve a higher level of solid fraction. 
As long as the vibration period $\tau = \nu^{-1}$ is longer than 
$\tau_c$, the packing has enough time  to relax fully to a more compact equilibrium state. 
But, when the  period $\tau$ is below  $\tau_c$, the relaxation is  incomplete so that 
 $\Delta \rho_1 < \Delta \rho_{max}$, where $\Delta \rho_{max}$ is the largest 
 compaction between two periods. 
  
It is expected that $\Delta \rho_1$ should follow the same scaling 
with the frequency as the displacement of 
the retaining wall, i.e.  $\Delta \rho_1 \propto \Delta \rho_{max} \ \nu^{-2}$. This is because 
the volume change $\Delta V$ is proportional to $\Delta x$. 
Hence, from Eq.  \ref{eq:eta1} and imposing the continuity at $\nu=\nu_c$, we get 
\begin{equation}
{\dot \eta}=
\left\{
\begin{array}{lll}
\frac{\Delta\rho_{max}}{\rho_0} \ \nu & \nu < \nu_c, \\
 \frac{\Delta \rho_{max}}{\rho_0} \nu_c^2 \ \nu^{-1} & \nu > \nu_c. 
\end{array}
\right.
\label{eq:eta1-2}
\end{equation}
Fig. \ref{fig18} shows this prediction together with the data points. 
We see that, although $\nu_c$ is the only fitting parameter,  
the compaction rate ${\dot \eta}$ is well adjusted by Eq. \ref{eq:eta1-2}. 
The prefactor $\Delta \rho_{max} / \rho_0$ is $ \simeq 0.005$,  
corresponding to   $\Delta \rho_{max} \simeq 0.0025$. 
In spite of the sharp transition at $\nu = \nu_c$, 
it is convenient to construct a single expression 
containing the correct behavior both at low and high 
frequencies. As in 2D for polygon packings \cite{Azema2006}, the
 following fitting form provides a good approximation 
as shown also in  Fig. \ref{fig18} (fitting form 2):
\begin{equation}
{\dot \eta} = \frac{\Delta\rho_{max}}{\rho_0} \frac{1 + e^{  - \left( \frac{\nu}{\nu_c} - 1  \right)^2 }}
{1 +  \left( \frac{\nu}{\nu_c}   \right)^2} \ \nu.
\label{eq:eta1-3}
\end{equation}

The characteristic time $\tau_c = 0.1$ s  is  of the same order of magnitude as 
the time required for one particle  
to fall down a distance equal to its diameter. 
Obviously, the above findings concern only short-time compaction 
($\Delta t < 1$ s). At longer times,  ${\dot \eta}$ declines with time, 
but the scaling with frequency according to Eq. \ref{eq:eta1-2} is 
expected to hold at each instant of evolution of the packing.


\section{Conclusion}
\label{sec8}

In this paper, the contact dynamics method was employed to simulate and analyze 
the dynamics of a system of polyhedral particles 
subjected to horizontal harmonic forcing of a retaining wall.  
Our system is devoid of elastic elements and, hence, 
the behavior is fully governed by particle rearrangements. 
Moreover, it involves  a jammed state separating passive (loading) 
and active (unloading) states. 
Dimensional analysis was used to scale the displacements 
with the frequency of oscillations. It was shown  that the data collapse 
by scaling the displacements by the inverse square of frequency. 
We also studied the scaling with confining force and 
particle weights. 

Granular ratcheting under horizontal vibrations was  investigated. 
During each vibration period a small compaction of the system 
occurs during unloading, i.e. upon sample extension, 
followed by  decompaction upon contraction. 
The compaction rate  increases linearly with frequency 
up to a characteristic frequency and then it declines in inverse 
proportion to frequency. 
The characteristic frequency was interpreted in terms of  relaxation time of the packing 
under its own weight during the unloading phase.  

The similarity of the phenomenology of  vibrational dynamics and compaction 
at short times in the 3D system of polyhedral particles with that of a 2D 
system of polygonal particles suggests that space dimensionality plays 
a minor role in granular dynamics. The characteristic times and compaction rates 
are slightly different but the scaling behavior and the functional dependence 
of the compaction rate with frequency are  the same. A comparison with 
spherical particles would be interesting in order to highlight the effect of 
particle shapes on these parameters. On the other hand, the characteristic time 
appears as a crucial parameter for the compaction rate and it merits 
further investigation as a function of various control parameters of the system.

The authors would like to thank specially F. Dubois for interesting discussions and help 
with the software LMGC90. This work was supported by a grant 
from the R\'egion Languedoc-Roussillon and the french railway company SNCF. 


\end{document}